\documentclass{article}
\usepackage[preprint, nonatbib]{nips_2018}

\usepackage[T1]{fontenc}% For Type-3 fonts
\usepackage{dsfont}
\usepackage{booktabs} % For formal tables

\usepackage[ruled]{algorithm2e} % For algorithms

\SetAlFnt{\small}
\SetAlCapFnt{\small}
\SetAlCapNameFnt{\small}
\SetAlCapHSkip{0pt}
\IncMargin{-\parindent}

\usepackage{amsmath}
\usepackage{comment}
\usepackage{mathtools}
\usepackage{courier}
\usepackage{xspace}
\usepackage{datetime}
\usepackage{adjustbox}
\usepackage{enumitem}
\usepackage{subfig}
\usepackage{url}
\usepackage[square,sort,comma,numbers]{natbib}

\newcommand{\sr}[1]{\textit{#1}\xspace}
\newcommand{\eytan}[1]{{\textcolor{purple}{\bf [*** EA: #1]}}}

\graphicspath{ {figs/} }

\begin{document}
% Title portion. Note the short title for running heads 
\title{Extracting Inter-community Conflicts in Reddit}  
\author{
  Srayan Datta \\
  University of Michigan\\
  \texttt{srayand@umich.edu} \\
  %% examples of more authors
  \And
  Eytan Adar \\
  University of Michigan\\
  \texttt{eadar@umich.edu}
}
\maketitle

\begin{abstract}
Anti-social behaviors in social media can happen both at user and community levels. While a great deal of attention is on the individual as an `aggressor,' the banning of entire Reddit subcommunities (i.e., subreddits) demonstrates that this is a multi-layer concern. Existing research on inter-community conflict has largely focused on specific subcommunities or ideological opponents. However, antagonistic behaviors may be more pervasive and integrate into the broader network. In this work, we study the landscape of conflicts among subreddits by deriving higher-level (community) behaviors from the way individuals are sanctioned and rewarded. By constructing a \textit{conflict network}, we characterize different patterns in subreddit-to-subreddit conflicts as well as communities of \textit{`co-targeted'} subreddits. By analyzing the dynamics of these interactions, we also observe that the \textit{conflict focus} shifts over time.

\end{abstract}

\textbf{keywords}: Inter-group conflict, Reddit, social network analysis, conflict network

\section{Introduction}
Anti-social behavior in social media is not solely an individual process. Communities can, and do, antagonize other groups with anti-social behaviors. Similarly, both individuals and communities can be sanctioned in reaction to this behavior. On Reddit, for example, individuals can be banned or otherwise be sanctioned (e.g., have their posts down-voted). Likewise, entire subreddits can also be sanctioned when multiple individuals use the community as a platform for generating conflict, in violation of general Reddit norms.  

Critically, the form of anti-social behavior at the community level can be quite varied. The ability to identify and coordinate with others means that actions considered anti-social for the individual can be expanded to group settings. A group can thus act anti-socially--- producing mass spamming and trolling, flame wars, griefing, baiting, brigading, fisking, crapflooding, shitposting, and trash talking---against both individuals and other subreddits~\cite{Kumar18, schneiderimpoliteness}. On Reddit, as in other discussion boards, the ability to create (multiple) accounts under any pseudonym can further exacerbate such behaviors.  Although a vast majority of users are generally norm-compliant, anonymity can lead to less inhibited behavior from users~\cite{Suler04}.  In aggregate, the result is an entire embedded network of subreddit-to-subreddit conflicts inside of the Reddit ecosystem. Research has found specific instances of these conflicts. Our goal is to inferentially identify the structure and dynamics of this \textit{community-to-community} conflict network at scale. 

%On Reddit, individual authors can subscribe to or post to subreddits -- sub-forums with specific ideologies, interests, and norms.

To achieve this, we address a number of challenges. First among them is the lack of explicit group membership. Group `membership' in Reddit, and systems like it, can be vague. While subscriptions are possible, individuals can display member-like behaviors by posting to subreddits they are not part of. Such behaviors---subscription and posting---are not, however, a clear indication of the individual's `social homes.' An individual can display both social and anti-social behaviors within the community via posting. Instead, what we seek is not simply to identify an individual's `home' but to further discriminate between \textit{social homes} and \textit{anti-social homes}. 

To achieve this separation, we apply a definition that extends Brunton's construct of spam: a community defines spam as (messaging) behavior that is not consistent with its rules and norms~\cite{brunton2012constitutive}. That is, we seek to separate \textit{norm-complaint} behaviors that indicate social membership and those that are \textit{norm-violating} (indicating an anti-social home). Rather than relying on a global definition of norms, we utilize the sanctioning and rewarding behavior of individual subreddits in response to norm violation and compliance respectively. An explicit measure we leverage is up- and down-voting on posted comments. While these are not the only kind of sanctions and rewards, they are (a) consistently used, and (b) can be aggregated both at the individual and community levels. As we demonstrate below, inference based on these lower level signals can help identify broader conflicts. 

A further appeal of the bottom-up approach is that the converse, top-down identification of sanctions at the subreddit level, does not provide a clear indication of conflict. First, this signal is sparse as the banning of subreddits remains rare. Except for explicit \textit{brigading}, which are (hard to detect) coordinated attacks on another subreddit, community-based anti-social behaviors may not result in a community being sanctioned. Second, even when a sanction is employed it may be due to other reasons than community-on-community attacks. For example, subreddits such as \sr{fatpeoplehate} (a fat-shaming subreddit) and \sr{europeannationalism} (a Nazi subreddit) have been banned but not necessarily due to any specific `attack' but rather non-compliance with Reddit-wide norms on hate speech. 

Our bottom-up inference is different in that we can identify pairs of social and anti-social homes and aggregate these to find conflicts. Specifically, we can find authors implicated in conflicts---which we call \textit{controversial authors}---by identifying those that have both social \textit{and} anti-social homes. From this, we can say that if multiple authors have a social-home in subreddit $A$ \textit{and} an anti-social home in subreddit $B$, then there is a directed conflict between $A$ and $B$. By finding aggregate patterns using all Reddit comments from 2016 (9.75 million unique users and 743 million comments), we can construct the subreddit \textit{conflict graph} at scale. Furthermore, we demonstrate how the directed edges in our graph can be weighted as a measure of \textit{conflict intensity}. The process of identifying conflict edges and their associated weights is complicated by the inherent noise in behaviors and high-variance of community sizes. A specific contribution of our work is the use of different aggregation and normalization techniques to more clearly identify the conflict graph.

Using this graph, we can determine not only the broad landscape of community-to-community conflicts but to answer specific questions as well: Which subreddits are most often instigators of conflict (versus targets)? Are conflicts reciprocal and are they proportional in intensity? Does `attacking' multiple subreddits imply broad misbehavior by members of that subreddit or the work of just a few individuals? Are certain subreddits targeted `together?' Do conflicts shift over time?

Briefly, we find that subreddit conflicts are often reciprocal, but the conflict intensity is weakly negatively correlated with the intensity of the `response.' We also find the larger subreddits are more likely to be involved in a large number of subreddit conflicts due to their size. However, our analysis of the fraction of users involved can isolate situations where both relative and absolute counts of involved authors are high. Additionally, we find different patterns of conflict based on intensity. For example, a single subreddit targeting many others may divide its attention, resulting in decreased intensity across the targets. On the other hand, we find anecdotal evidence that subreddits which act as social homes to many controversial authors and have high average conflict intensity against other subreddits often display communal misbehavior. Because of the longitudinal nature of our data, we are also able to perform a dynamic analysis to isolate temporal patterns in the conflict graph. We find, for example, that subreddits that conflict with multiple other subreddits change their main focus over time.

Our specific contributions are mapping the static and dynamic subreddit conflict networks across Reddit. We identify group membership and define the concept of social and anti-social homes as a way of defining conflicts. By analyzing the different static and temporal patterns in subreddit conflicts, we provide evidence for mechanisms that can identify communal misbehavior. We provide a baseline for quantifying conflicts in Reddit and other social networks with `noisy' community structure and where individuals can behave (and misbehave) in a communal fashion. Our work has implications in identifying community features which can be used to automatically monitor community (mis)behavior in such social networks as an early warning system.    

%This presents the first challenge in building a map of subreddit conflicts: not all anti-social behaviors are conflicts, and not all conflicts are indicated by banning. 

%Rather than utilizing the broad norms of Reddit, we can identify behaviors that are explicitly rejected by a community and can thus be considered anti-social/antagonistic form the lens of that group. This is c as being defined by the targeted community.
%Posting behavior may be a better indicator, but not all posts may be `agreeable' to the subreddit. In fact, a post can be norm-compliant or norm-violating. To map individuals to communities, we introduce the idea of an author's \textit{social home} and \textit{anti-social home}. These roughly map to subreddits where an author's posts are viewed positively (norm-compliant) or negatively (norm-violating). This perspective allows for contextualizing conflicts. 

%A second challenge in building the conflict graph is in identifying community membership. That is, before we can aggregate individual behaviors, we need to understand who is part of that community. Like other forums, Reddit's flexible definition of `membership' makes this difficult. 

\section{Related Work}
\subsection{Conflicts in Social Media}
Undesirable behavior in online communities are widely studied in social media research. Qualitative analysis often focuses on identifying and characterizing different types of inappropriate online behaviour or provides case studies in different forums. Analysis of Usenet news, for example, helped explore identity and deception~\cite{donath1999identity}. General anti-social behaviors~\cite{hardaker2010trolling} can also manifest in `site-specific' ways as in the trolling and vandals on Wikipedia~\cite{Shachaf2010}, or griefing and combative strategies in Second Life~\cite{Chesney2009}.  

Predicting trolls and other anti-social behavior is another well explored research area. For example, researchers have studied the connection between trolling and negative mood which provided evidence for a `feedback' loop that contributes to further trolling~\cite{Cheng17anyone}. In the context of \textit{prediction}, several studies focused on detecting certain anti-social behaviors on specific sites (e.g., vandalism in Wikipedia~\cite{Adler2011,Kumar2015}). Others have attempted to predict both anti-social behaviors or sanctions. Examples of the former include finding sockpuppets (same user using multiple accounts) on discussion sites~\cite{Kumar2017sockpuppet} and Twitter~\cite{GalnGarca2013SupervisedML}. Within Reddit, Kumar et al.~\cite{Kumar18} studied controversial hyperlink cross-postings between subreddits to identify community conflict. Examples of the latter (sanction prediction) include future banning based on comments~\cite{Cheng15antisocial} and using abusive content on one forum to predict abuse on others~\cite{Chandrasekharan2017}.  The bulk of research has emphasized the behavior of individuals rather than inter-community anti-social behavior (rare exceptions emphasized specific types of anti-social behavior). While we draw upon this literature to understand individual trolling, our focus is on a broad definition of higher-order inter-community conflicts. That is, our aim is to identify inter-community conflict (rather than individual-on-individual or individual-on-community) by developing behavioral mapping mechanisms in the context of the broader network. 

\subsection{Trolling in Reddit}
Individual trolling in Reddit is predominantly studied through content analysis (e.g., ~\cite{merritt2012analysis}). A key result for Reddit has been comparing the differences between a smaller number of communities in terms of trolling behavior. For example, Schneider performed a contrastive study on intercultural variation of trolling by two subreddits, \sr{ShitRedditSays} and \sr{MensRights}~\cite{schneiderimpoliteness}. Most related to our work is the study by Kumar et al.~\cite{Kumar18} which found that very few subreddits are responsible for the majority of conflicts. This has implications to the conflict graphs we construct in that we may expect key conflict `nodes.'

More recently, there has been research on interventions (e.g., banning) to combat anti-social behavior. For example, Chandrasekharan et al.~\cite{Chandrasekharan2017ban} studied the effect of banning two particular subreddits, \sr{fatpeoplehate} and \sr{CoonTown}, to combat hate-speech. However, this work does not elaborate on subreddit-to-subreddit relations before or after the ban. Subreddit relations are discussed from an ideological frame by identifying subreddits which discuss the same topic from different point of views~\cite{Datta2017z2}. However, this approach does not capture conflict explicitly.    

\subsection{Signed Social Networks}
We analyze subreddit conflicts by creating a subreddit conflict graph, which can be viewed as a \textit{signed graph} (where all the edges are marked negative). Use of signed graphs for trolling detection is uncommon but has been explored in past research. Kunegis et al.~\cite{Kunegis2009} predicted trolls and negative links in Slashdot (a technological news website and forum where users are able to tag other users as `friend' or `foe'). Multiple studies~\cite{Wu2016,Shahriari2014} proposed models to rank nodes in signed social networks. Signed networks incorporate both positive and negative edges. In our case, it is difficult to make claims about positive relations in the conflict graph. Because most individuals are norm-compliant, edges constructed between two \textit{social} homes may be an artifact of authors being largely norm-compliant and simply reflect correlated interests. In contrast, an author that displays both norm-compliant and norm-violating behaviors provides a better indication of likely conflict.

\begin{comment}
- Trolling in social media\\
- Trolling in Reddit\\
- Subreddit wars\\

\eytan{I would suggest trying to get through this sooner rather than later, it will help you frame some of your findings and possibly identify key RQs}
\end{comment}
\section{Dataset}
In our research, we focus on Reddit (\url{www.reddit.com}) both due to its similarity (in features) to many other discussion boards and its vast scale. Reddit is a social aggregator and discussion forum for millions of individuals who regularly post news, video, images or text and discuss them in different comment threads. Reddit divides itself into focused subreddits which discuss particular topics or perform specific kinds of social aggregation (e.g., image or video sharing). In each subreddit, user-submitted content is referred to as a Reddit \textit{post}, and a post is discussed via different dedicated \textit{comment threads}. Each post and comment can be rewarded or sanctioned by other Reddit users using upvotes and downvotes. Within each subreddit, comments can also be flagged or moderated by subreddit-specific moderators. Individuals who behave inappropriately can also be banned from specific subreddits.

For the analysis presented here, we used all publicly available Reddit comments from 2016. This was a subset of the multi-year Reddit data (posts, authors, comments, etc.) compiled by Baumgartner~\cite{pushshift}. We specifically mined commenting behavior (rather than posting) for building conflict graphs. Comments are much more prevalent than posts, and anti-social behavior in Reddit often involves inflammatory comments rather than posts. For each comment, we make use of the following metadata: author of the comment, which subreddit the comment was posted on and how many upvotes and downvotes the comment received. We found that there are 9,752,017 unique authors who commented at least once in Reddit in our sample. Though largely a `human population,' bots can also be programmed to generate comments. Of the 9.7M authors, 1,166,315 were `highly active,' posting more than 100 comments throughout the year. On average, a Reddit user posted in 7.2 subreddits and commented 76.2 times in 2016. As may be expected, most Reddit users are pro-social. In 2016, we find that 79.2\% of authors (across all of Reddit) have at least 90\% of their comments upvoted.  
      
\section{Identifying Inter-community Conflicts}
To define the \textit{conflict graph} between subreddits we need first to identify edges that capture community-on-community `attacks.' We would like these edges to be directed (as not all conflict is reciprocated) and weighted (to indicate the strength of the conflict). Our goal is not to only identify `passive' ideological opposition but also behaviors where one subreddit actively engages with the other.

This distinction is important as there are instances where two subreddits are discussing the same topic through different ideologies (as determined through text analysis), but have very low author overlap. For example, the \sr{askscience} (discussion forum for science-related topics) and \sr{theworldisflat} (forum for scientific evidence that the world is flat) could be considered to be ideologically opposed~\cite{Datta2017z2}. However, there are very few authors who post in both subreddits, meaning there is no engagement and no `conflict' by our definition. These individuals do not agree but largely leave each other alone. 

Instead, we focus on identifying individuals that post to multiple subreddits and behave \textit{differently} depending on the subreddit. In our model, behaviors, such as commenting, can be norm-compliant or norm-violating. Norm-compliant are those behaviors that the community finds agreeable in that they are consistent both with the way behaviors (e.g., message posting) should be done and/or the content of the message itself.  Norm-violating are those behaviors that are disagreeable in the same way (how they're posted or what is in them). Norm-violating behavior can include traditionally anti-social behaviors:  flame wars, griefing, spamming, trolling, baiter, brigading, baiting, fisking, crapflooding, shitposting, and trash talking. This, again, is consistent with Brunton's spam definition~\cite{brunton2012constitutive}. The appeal of this localized definition of spam is that each community can assert what they consider social or anti-social behavior (i.e., norm-compliant and norm-violating) and can make local decisions to reward or sanction such behaviors respectively.

Our inferential goal is to operationalize social and anti-social behavior by leveraging reward and sanction behaviors as indicators. For this purpose, we use up- and down-votes.  Obviously, not all compliant behaviors are rewarded through up-votes, nor are all norm-violating sanctioned through down-votes (banning being a notable alternative). Other metrics for norm-violation may include identifying banned users or users whose comments are regularly removed by moderators. Unfortunately, such data is not readily available (removed comments and authors will be missing from the dataset). Posts can also be marked as `controversial' to signal undesirable behavior, but these are not always anti-social per se. Additionally, both banning and controversial post `tagging' may not be reliably imposed. Upvoting and downvoting, however, are specifically part of the incentive structure for Reddit and are both uniformly applied and ubiquitous.

\begin{figure}[h]
 \centering
 \includegraphics[scale=0.65]{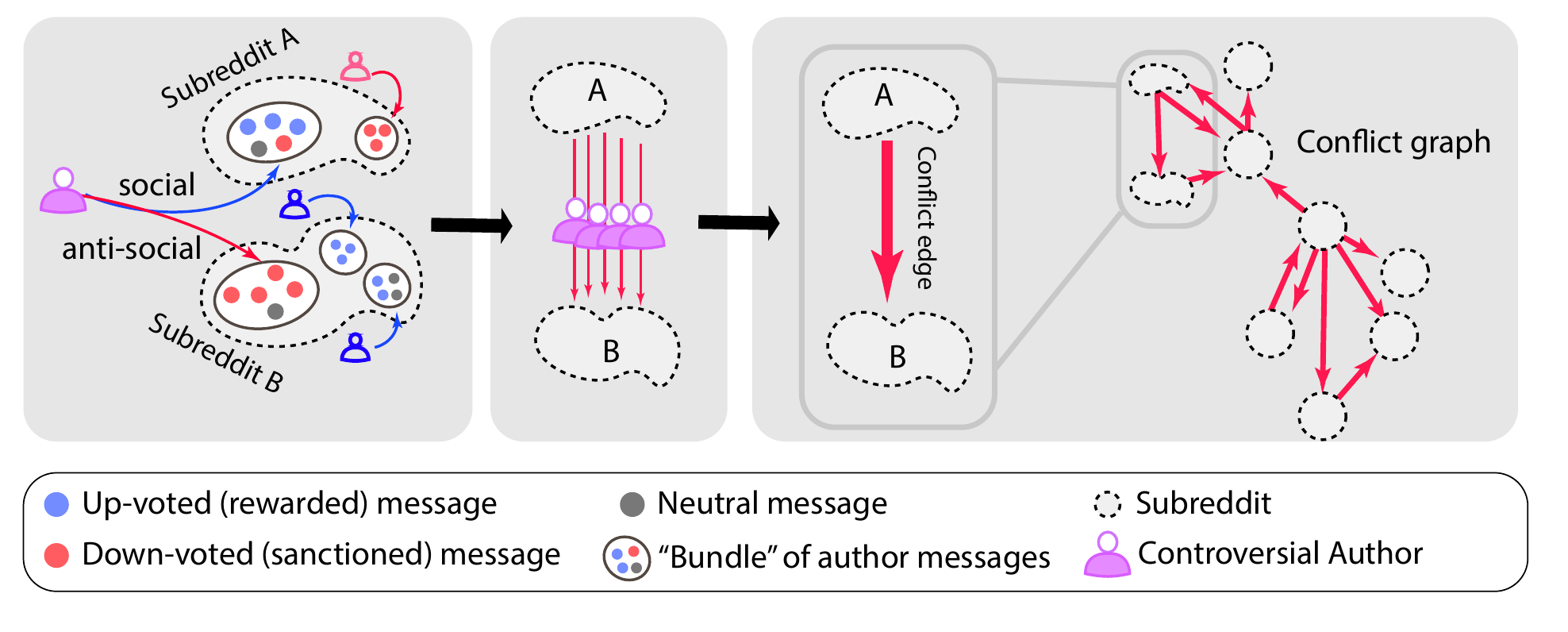}
 \caption{General methodology for identifying conflicts and creating the conflict graph.}
\label{fig:conflictgraph_vis}
\end{figure}

An individual who reliably produces enough \textit{measurable norm-compliant behavior} (e.g., many upvoted messages) can be said to have a \textit{social home} in that community. Likewise, an individual that produces a substantial amount of \textit{measurable norm-violating behavior} (i.e., many downvotes) is said to have an \textit{anti-social home} in that community. An individual can have multiple social and multiple anti-social homes. Because our goal is to find conflict \textit{edges} we do not consider authors that are \textit{only} social or \textit{only} anti-social. Those who are globally norm-violating (e.g., spammers, malicious bots, etc.) and are negatively treated in all subreddits in which they post are removed from consideration. Figure~\ref{fig:conflictgraph_vis} (left) illustrates this idea.

A second key aspect in building the conflict graph is in aggregation. One \textit{particular} individual may have a social home and an anti-social home. However, from the single example, we can not infer that the other members of that person's social home would endorse the messages the person is posting to the other subreddit. Instead, we look for signals in the aggregate. If there are many individuals, who cross-post to two subreddits--- where one subreddit is clearly a social home, and the other is clearly an anti-social home---we infer that a conflict exists. This conflict need not be reciprocated, but as we show below, it often is.

We can roughly quantify the anti-social behavior of a user within a subreddit if he/she has more downvoted comments compared to upvoted comments. Note that a single comment can have multiple upvotes and downvotes. Reddit automatically upvotes a user's own comment (all comments in Reddit start with one upvote). We consider the user's upvote as a `baseline' as we assume the author views his or her own comment positively\footnote{The algorithms described in this paper can be applied with or without a individuals' personal upvotes. However, we note that exclusion of this number may slightly change the descriptive statistics we report.}. We say a comment is \textit{downvoted} (in aggregate) if the total number of downvotes for the comments exceeds upvotes, and \textit{upvoted} when upvotes exceed downvotes. Similarly, we determine that a user has shown social behavior if they have more upvoted comments (rewarded, norm-compliant) compared to downvoted ones (sanctioned, norm-violating) within a subreddit. 

To distinguish between an author's `home' and simply a `drive-by' comment, we enforce a threshold (we call this \textit{significant presence}) of more than ten comments in the subreddits over the course of the year (2016). This threshold also ensures that we can observe enough up and down votes for any particular author. Additionally, new authors in a subreddit might break some unfamiliar rules, and receive downvotes initially. Our threshold gives sufficient data to determine if they `learned.' We also enforce that the user has more than 100 total posts in 2016, which ensures that they have an overall significant presence on Reddit.

As authors were automatically assigned to \textit{default subreddits} (\sr{AskReddit}, \sr{news}, \sr{worldnews}, \sr{pics}, \sr{videos}), many Reddit authors began by posting in these groups~\footnote{Though this does not impact our analysis (for 2016 data), we note that default subscription was replaced in 2017 with a dynamic popular subreddit homepage.}. Norms (and norm-compliance) in these subreddits may be significantly different from rest of the subreddits. Using our definition of social homes, a large number of users have at least some default subreddit as their social home or anti-social home just because they started by posting in these forums. For this reason, we exclude default subreddits from our analysis.  

\subsection{Controversial Authors}
We denote an author with at least one social and one anti-social home as a \textit{controversial author} (the purple figures in Figure~\ref{fig:conflictgraph_vis}). In 2016, 1,166,315 authors had more than total 100 comments over the year. After filtering for significant presence in subreddits, we found 23,409 controversial authors. This indicates that only about 2\% of the more prolific Reddit users fall into this category. Among the controversial authors, 82\% have only a \textit{single} anti-social home. The vast majority (92.5\%) of controversial authors have \textit{more} social than anti-social homes. This indicates that these authors differ from the conventional idea of a ``troll'' who misbehaves in every forum they participate in. This also means that a typical controversial author focuses his/her `misbehavior' on a small number (usually 1) subreddits. This result is consistent with Reddit users being loyal (in posting) to a small set of subreddits~\cite{Hamilton2017}.

In aggregate, if there are many controversial authors that have a social home in subreddit $A$ and anti-social home in subreddit $B$, we view this to be a directed conflict from $A$ against $B$. We call this a \textit{conflict edge}. The sum of all these edges, after some additional filtering, captures the \textit{conflict graph} (Figure~\ref{fig:conflictgraph_vis}, right).

Our approach has the benefit that aggregation can eliminate various types of noise. While upvoting/downvoting is noisy at the level of any particular message, aggregation at the author level allows us to look for \textit{consistent} behaviors (i.e., are messages from an author always rewarded in one place and sanctioned in another?). Noise at the level of a \textit{particular} controversial author is similarly mitigated by aggregation (are there \textit{multiple} individuals being rewarded in one place and sanctioned in the other?).

\section{The Subreddit Conflict Graph}
\subsection{Constructing the Conflict Graph}
To construct the conflict graph, we apply the following strategy. If $k$ authors have a social-home in subreddit $A$, and an anti-social home in subreddit $B$, we can create a weighted directed \textit{conflict edge} from $A$ to $B$. If we create these edges for all subreddit pairs, we have a graph of antagonistic subreddit relations. Weights for these edges must be normalized as different subreddits have a different number of users. Thus, a raw author count (i.e., common authors with a social home in $A$ and anti-social home in $B$) is misleading. Larger subreddits would dominate in weights as more authors often means more controversial authors.  For convenience, we refer to the `source' of the edge as the \textit{instigating} subreddit and the `destination' as the \textit{targeted} subreddit.

We normalize the raw controversial author counts by the number of common authors in both subreddits. Furthermore, for each subreddit pair, we require that there are at least five controversial authors between them to ensure that we are not misidentifying a conflict due to very few controversial authors (i.e. if there is $k_1$ authors with social home in subreddit $A$ and anti-social home in subreddit $B$, and $k_2$ authors with social home in subreddit $B$ and anti-social home in subreddit $A$, $k_1+k_2$ must be at least five). We emphasize that the weight, direction, or even existence, of an edge from subreddit $A$ to $B$, is very different from an edge from $B$ to $A$. 

\subsection{Eliminating Edges Present due to Chance}
While defining conflict between a pair of subreddits, we need to make sure that users are not perceived negatively in the attacked subreddit by chance. For two subreddits $A$ and $B$ with $n_{common}$ common users and $n_{actual}$ users perceived positively in subreddit $A$ but negatively in subreddit $B$ (we only consider users who posted more than 10 times in both subreddits), we calculate the number of users who can be perceived negatively in subreddit $B$ by chance. First, we define an empirical multinomial distribution of comment types for subreddit $B$, i.e., we calculate the probabilities of a random comment in subreddit $B$ being positive (upvoted), negative (downvoted) or neutral. To create this multinomial distribution, we only use comments from users who posted more than ten times in subreddit $B$ as these are the users we consider when declaring controversial authors. For a common user $i$, if $i$ posted $n_i$ times in subreddit $B$, we sample $n_i$ comments from the probability distribution and calculate if he/she is perceived negatively in the sample. We sample all common users and count the total number of users perceived negatively in subreddit $B$. We repeat this experiment 30 times to create a sampling distribution of the expected number of negatively perceived users and calculate the $z$-score of $n_{actual}$ using this sampling distribution. We only retain conflicts from $A$ to $B$, where this $z$-score is greater than 3, i.e., the number of users perceived negatively in the attacked subreddit is significantly higher than the number expected from random chance. The final set of subreddits (nodes) and associated edges are the \textit{conflict graph}. %(Figure~\ref{fig:hairball}). 

\subsection{Conflict Graph Properties}
The final subreddit conflict graph for 2016 consists of 746 nodes and 11,768 edges. This is a small fraction of active subreddits in 2016 (around 76,000) which is, in part, due to the low amount of `multi-community posting' on Reddit~\cite{Hamilton2017} (i.e., very few authors regularly post to more than one `home' community). As we require multi-community posts to create an edge, the result is that many subreddits are `free floating' and are removed from consideration. Of the 746, nine were banned sometime between the end of 2016 and April of 2018: \sr{PublicHealthWatch} (a subreddit dedicated to documenting the `health hazards' of, among others, LGBTQ groups), \sr{altright}, \sr{Incels} (involuntary celibate), \sr{WhiteRights}, \sr{european}, \sr{uncensorednews}, \sr{europeannationalism}, \sr{DarkNetMarkets} and \sr{SanctionedSuicide}. An additional six became `private' (requiring moderator approval to join and post), which includes a couple of controversial subreddits: \sr{Mr\_Trump} and \sr{ForeverUnwanted}.

\begin{figure}[h]
 \centering
 \includegraphics[scale=1]{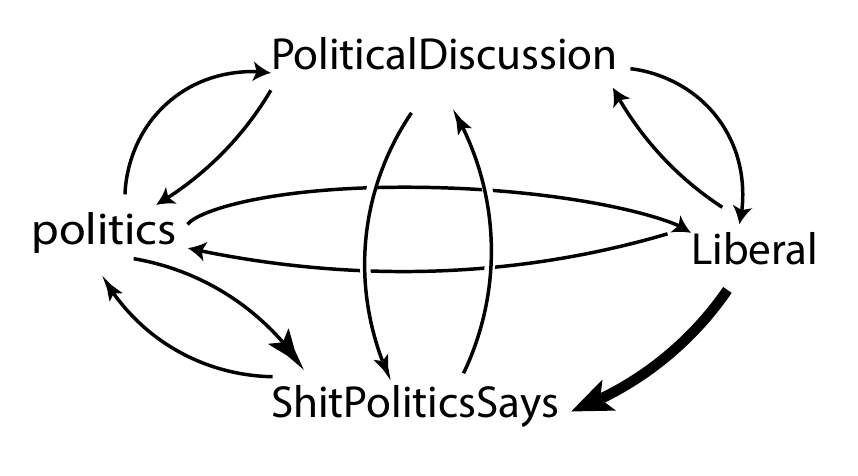}
 \caption{Ego network for the subreddit \sr{Liberal}. Thicker edges denote higher conflict intensity.}
\label{fig:liberal}
\end{figure}

The conflict graph consists of 5 components, with the giant component containing 734 nodes. The next largest component consists of only 6 nodes representing different sports streaming subreddits (\sr{nflstreams}, \sr{nbastreams}, \sr{soccerstreams} etc.). Through manual coding of subreddits we identify the following high-level categories: political subreddits (e.g. \sr{politics}, \sr{The\_Donald}, \sr{svenskpolitik}) discussion subreddits, video game subreddits (e.g. \sr{Overwatch}, \sr{pokemongo}), sports fan clubs, location-focused subreddits (e.g. \sr{canada}, \sr{Seattle}, \sr{Michigan}, \sr{Atlanta}), subreddits for marginalized groups (e.g. \sr{atheism}, \sr{DebateReligion}, \sr{TrollXChromosomes}, \sr{lgbt}, \sr{BlackPeopleTwitter}), *porn subreddits (these are image sharing subreddits with their name ending with porn, they are not pornographic in nature -- e.g. \sr{MapPorn}, \sr{HistoryPorn}) and NSFW subreddits (e.g. \sr{nsfw}, \sr{NSFW\_GIF}). Because of our use of 2016 data and the associated (and contentious) election, political subreddits are heavily represented in the conflict graph. Figure~\ref{fig:liberal} shows the ego network for the subreddit \sr{Liberal} in the conflict graph. 

Edge weights in the conflict graph are often low. On average, only 3.57\% (median is 1.70\%) of authors in the `conflict source' subreddit (i.e., their social home) post to the target subreddit (i.e., their anti-social home). There are, however, edges with extremely high weights. The highest edge weight in our data is 85.71\% from \sr{The\_Donald} to \sr{PanicHistory}. However, in this case, this is due to the disproportionate difference in size of the two (they share only seven common authors). Thus, a high conflict intensity does not necessarily mean that a large fraction of originating subreddit users are antagonistic to the target subreddit. Nonetheless, it does point to the fact that larger subreddits with many controversial authors can overwhelm a smaller subreddit. The high edge-weight here indicates the degree to which this happens. Using the subreddit conflict graph, we can isolate the main source and targets of conflicts and understand where conflicts are one-sided or mutual.

\begin{figure}[h]
 \centering
 \includegraphics[scale=0.55]{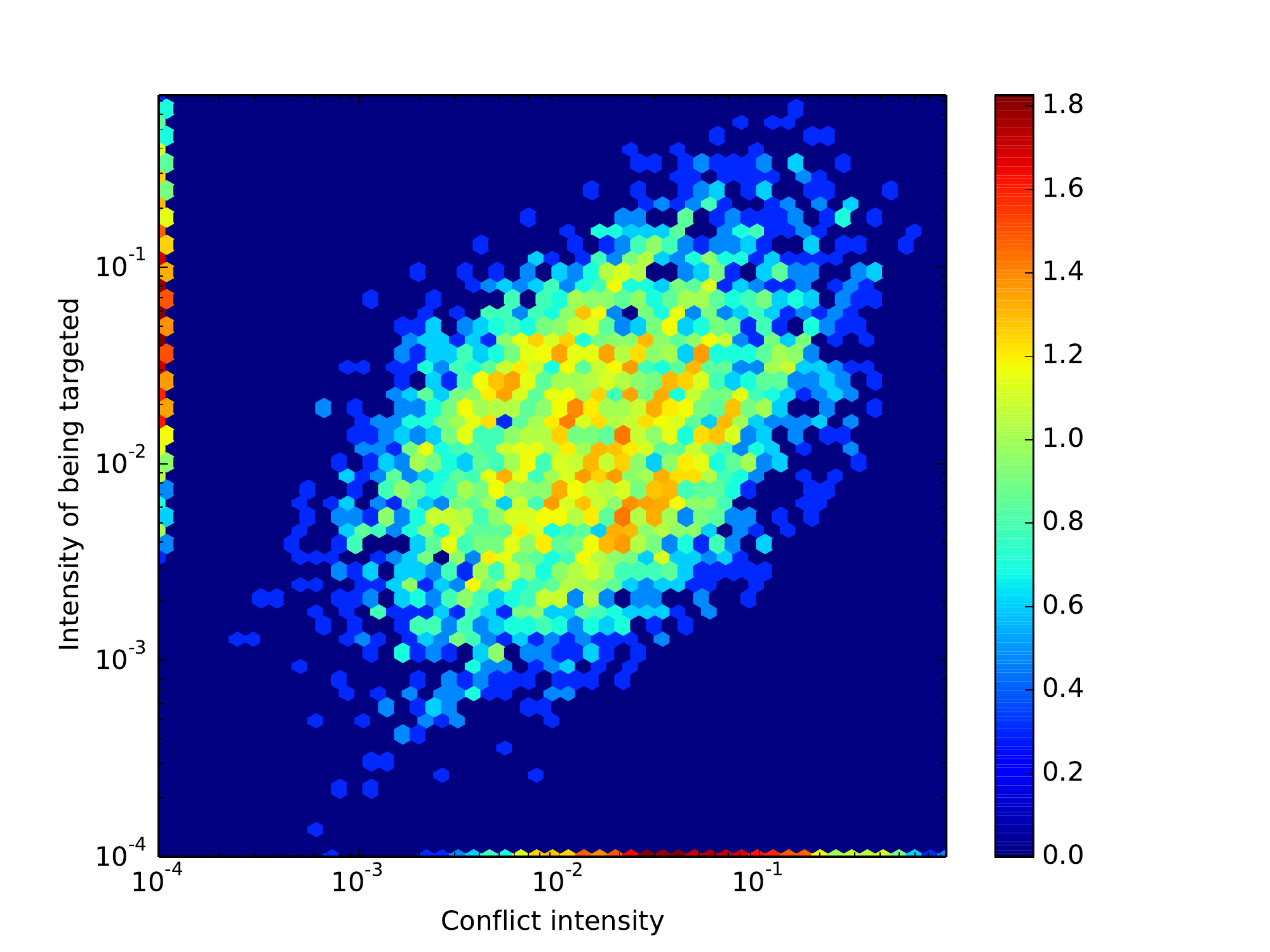}
 \caption{Conflict intensity vs intensity of reciprocation in the subreddit conflict graph (log-scale). Un-reciprocated edges appear at the bottom and left edge.}
\label{fig:attack}
\end{figure}

\textbf{Are conflicts reciprocal?} We find that if a conflict edge exists between subreddit $A$ and $B$, in 77.2\% cases the inverse edge will \textit{exist}. Calculating the Spearman correlation between conflict intensities of pairs of reciprocated edges, $\rho(5126)=-0.111, p < 0.0001$, we observe a weak (but significant) negative relationship. Figure~\ref{fig:attack} depicts the outgoing conflict (source) intensity versus incoming conflict (the conflict target) intensity. This indicates that a targeted subreddit usually reciprocates, but the intensity is usually not proportional. 

\textbf{Which subreddits are most targeted in 2016?} The indegree of a subreddit roughly indicates the number of other subreddits targeting it. The weighted sum of these edges (weighted indegree) corresponds to the intensity. The top 10 most targeted subreddits by indegree are \sr{politics}, \sr{SubredditDrama}, \sr{AdviceAnimals}, \sr{EnoughTrumpSpam}, \sr{atheism}, \sr{SandersForPresident}, \sr{The\_Donald}, \sr{PoliticalDiscussion}, \sr{technology} and \sr{KotakuInAction} respectively. However, when we order subreddits by \textit{total incoming conflict intensity} (see Table~\ref{table:wt_indeg}) the list is somewhat different. In both lists, we observe that the most targeted subreddits are social and political discussion forums as well as forums that discuss Reddit itself. The heavy presence of political forums can be attributed to the 2016 US presidential election. We try to deduce if, in general, the most targeted subreddits by degree are also the most targeted subreddits by average incoming intensity (total intensity/number of sources) and vice versa. When contrasting indegree to average intensity for subreddits that are targeted by at least one subreddit (we have 673 such subreddits), we find a weak positive correlation with Spearman $\rho(673) = 0.242, p < 0.0001$. A subreddit targeted by many subreddits is not necessarily targeted with high intensity. Conversely, subreddits targeted by only a few others can nonetheless be targeted with high intensity. 

\begin{table}[htb] 
\centering
\begin{adjustbox}{max width=\columnwidth}
\begin{tabular}{|c|c|c|}
\hline
\textbf{Subreddit} & \textbf{Indegree} & \textbf{Weighted indegree} \\
\hline
\sr{SubredditDrama} &  272  &  19.51 \\
\hline
\sr{EnoughTrumpSpam} &  217  &  13.25 \\
\hline
\sr{BestOfOutrageCulture} &  46  &  10.59 \\
\hline
\sr{ShitPoliticsSays} &  48  &  10.29 \\
\hline
\sr{Enough\_Sanders\_Spam} &  48  &  9.80 \\
\hline
\sr{sweden} &  81  &  9.62 \\
\hline
\sr{KotakuInAction} &  168  &  9.19 \\
\hline
\sr{ShitAmericansSay} &  94  &  8.83 \\
\hline
\sr{PoliticalDiscussion} &  185  &  8.08 \\
\hline
\sr{vegan} &  71  &  7.26 \\
\hline
\end{tabular}
\end{adjustbox}
\caption{Top 10 most targeted subreddits by total incoming intensity.}\label{table:wt_indeg} 
\end{table}

\textbf{Which are the most conflict `instigating' subreddits in 2016?} By using the conflict graphs outdegree (weighted or not) we can similarly find the largest conflict sources. The top-10 subreddits ranked by outdegree are \sr{politics}, \sr{AdviceAnimals}, \sr{The\_Donald}, \sr{SandersForPresident}, \sr{WTF}, \sr{technology}, \sr{atheism}, \sr{SubredditDrama}, \sr{EnoughTrumpSpam} and \sr{PoliticalDiscussion}. 

When ordered by total intensity, the top-10 list changes to include more news, politics, and controversy focused subreddits (Table~\ref{table:wt_outdeg}). This list also includes a now banned subreddit (\sr{uncensorednews}). However, we observe that most of these subreddits have low average conflict intensity (i.e., intensity per edge is low). If we order by average intensity (Table~\ref{table:avg_outdeg}), we find that subreddits targeting very few others (usually 1 or 2 subreddits) show up at top spots. However, we find that the subreddits at the first, third and ninth position of this list (\sr{europeannationalism}, a Nazi subreddit, \sr{PublicHealthWatch}, an anti-LGBT subreddit and \sr{WhiteRights}) are banned by Reddit. A controversial now private subreddit (\sr{ForeverUnwanted}) also appears in this list. This may have implications for identifying problematic subreddits. 

As before, we can check if the subreddits most often at the source of a conflict (by outdegree) are also the most instigating (by average conflict intensity). Using 719 `source' subreddits in our conflict graph, we find a weak positive correlation between the number of targeted subreddits and the average outgoing conflict intensity (Spearman $\rho(719) = 0.189, p < 0.0001$), which falls in line with our previous discussion. 

\begin{table}[htb]
\centering
\begin{adjustbox}{max width=\columnwidth}
\begin{tabular}{|c|c|c|}
\hline
\textbf{Subreddit} & \textbf{Outdegree} & \textbf{Weighted outdegree}  \\
\hline
\sr{The\_Donald} &  260  &  17.75 \\
\hline
\sr{politics} &  542  &  10.15 \\
\hline
\sr{conspiracy} &  141  &  7.39 \\
\hline
\sr{KotakuInAction} &  152  &  7.35 \\
\hline
\sr{uncensorednews} &  113  &  7.14 \\
\hline
\sr{AdviceAnimals} &  268  &  6.76 \\
\hline
\sr{SandersForPresident} &  211  &  6.62 \\
\hline
\sr{CringeAnarchy} &  148  &  6.12 \\
\hline
\sr{ImGoingToHellForThis} &  114  &  5.99 \\
\hline
\sr{Libertarian} &  92  &  5.86 \\
\hline
\end{tabular}
\end{adjustbox}
\caption{Top 10 subreddits (conflict source) ranked by total conflict intensity.}\label{table:wt_outdeg} 
\end{table}

\begin{table}[htb] 
\centering
\begin{adjustbox}{max width=\columnwidth}
\begin{tabular}{|c|c|c|}
\hline
\textbf{Subreddit} & \textbf{Outdegree} & \textbf{Average outdegree}  \\
\hline
\sr{europeannationalism} &  1  &  0.75 \\
\hline
\sr{OffensiveSpeech} &  1  &  0.62 \\
\hline
\sr{PublicHealthWatch} &  1  &  0.62 \\
\hline
\sr{askMRP} &  1  &  0.57 \\
\hline
\sr{ForeverUnwanted} &  2  &  0.50 \\
\hline
\sr{theworldisflat} &  1  &  0.43 \\
\hline
\sr{FULLCOMMUNISM} &  2  &  0.42 \\
\hline
\sr{marriedredpill} &  1  &  0.38 \\
\hline
\sr{WhiteRights} &  2  &  0.34 \\
\hline
\sr{SargonofAkkad} &  2  &  0.33 \\
\hline
\end{tabular}
\end{adjustbox}
\caption{Top 10 subreddits (conflict source) ranked by average conflict intensity.}\label{table:avg_outdeg} 
\end{table}

\textbf{Do larger subreddits get involved in more conflicts due to their size?} We find that larger subreddits are more likely to get involved in both incoming and outgoing conflicts. Using number of unique authors who posted more than 10 times in 2016 in the subreddit as a measure of subreddit size, we find moderate positive correlation between both size and number of incoming conflicts (Spearman $\rho(673) = 0.403, p < 0.0001$), and size and outgoing conflicts (Spearman $\rho(719) = 0.457, p < 0.0001$). However, taking conflict intensities into account, we find size and average incoming conflict intensity is moderately negatively correlated (Spearman $\rho(673) = -0.594, p < 0.0001$). Similarly, size and average outgoing conflict intensity is also weakly negatively correlated (Spearman $\rho(719) = -0.222, p < 0.0001$). This tells us that subreddits with larger size are more likely to be involved in conflicts just because there are more authors commenting in them, but average conflict intensity is not indicative of subreddit size. 

\subsubsection{Node properties}
Edge weights alone do not tell us if controversial authors are particularly prevalent in a specific subreddit. Rather, it only indicates the fraction of common users who are sanctioned (norm-violating) in the target subreddits. However, these common users might represent only a small fraction of users of a subreddit. This is especially possible for the larger subreddits. 
To determine which subreddits are the social home for \textit{many} controversial authors, we use three additional metrics: \textit{con\_author\_percent} is the percentage of controversial authors who make their social home in a subreddit relative to the number of authors who posted in that subreddit (more than 10 times in the year); \textit{avg\_subs\_targeted} and \textit{median\_subs\_targeted} are the average and median of number of subreddits that these controversial authors `target.' These numbers can tell us (a) what fraction of a subreddit are engaged in conflict, and (b) are they engaging in broad (across many subreddits) or focused conflicts. We limit our study to subreddits with at least 20 controversial authors who have a social home on that subreddit (overall, we find 698 subreddits meet this criterion). Removing smaller subreddits minimally affects the top-10 subreddits (see Table~\ref{table:con_percent}) by con\_author\_percent (only \sr{theworldisflat}, with 11 controversial authors, is removed from the list). We refrain from listing one pornographic subreddit in the table at rank 8.

\begin{table}[htb] 
\centering
\begin{adjustbox}{max width=\columnwidth}
\begin{tabular}{|c|c|c|c|}
\hline
\textbf{Subreddit} & \textbf{Con\_author\_percent} & \textbf{Average} &\textbf{Median}\\
\hline
\sr{PublicHealthWatch} & 35.25 & 2.44 & 2.0\\
\hline
\sr{OffensiveSpeech} & 34.78 & 2.86 & 2.0\\
\hline
\sr{WhiteRights} & 32.60 & 3.19 & 2.0\\
\hline
\sr{ThanksObama} & 32.59 & 2.34 & 1.5\\
\hline
\sr{europeannationalism} & 32.43 & 2.77 & 2.0\\
\hline
\sr{subredditcancer} & 27.51 & 2.33 & 2.0\\
\hline
\sr{subredditoftheday} & 24.90 & 1.94 & 1.0\\
\hline
\sr{POLITIC} & 23.94 & 1.91 & 1.0\\
\hline
\sr{undelete} & 23.04& 2.13 & 1.0\\
\hline
\sr{SRSsucks} & 22.18 & 2.07 & 1.0\\
\hline
\end{tabular}
\end{adjustbox}
\caption{Top 10 subreddits with highest percentage of positively perceived controversial authors (with at least 20). The average and median columns correspond to \textit{avg\_subs\_targeted} and \textit{median\_subs\_targeted}  respectively.}\label{table:con_percent} 
\end{table}

Most subreddits in top 10 list are either political forums or somewhat controversial in nature. To lend further credence to this measure, \sr{PublicHealthWatch} (an anti-LGBT subreddit, rank 1), \sr{WhiteRights} (rank 3) and \sr{europeannationalism} (a Nazi subreddit, rank 5) score highly with our metric and were recently banned by Reddit. It is also important to note that most controversial authors have only one anti-social home. Thus in almost all cases, median\_subs\_targeted is 1. The only exceptions are the first six subreddits in the table~\ref{table:con_percent}, \sr{The\_Farage} (median is 2) and \sr{sjwhate} (median is 2). Note that all banned subreddits shown in this table have a median of 2. The median con\_author\_percent for all 698 subreddits is 4.09\%, and the lowest is 0.36\%. It is worth noting that, the three banned subreddits in this list targeted only one or two subreddit each but with very high conflict intensity (e.g., \sr{europeannationalism} attacked \sr{AgainstHateSubreddits} with conflict intensity of 0.75). All three subreddits also show up in the list of most conflict-source subreddits by average conflict intensity. This shows that a subreddit does not have to target multiple other subreddits to be problematic. 

Compared to the top-10 subreddits by con\_author\_percent, large political subreddits in the most instigating subreddit list (conflict source) had a lower percentage of controversial authors who engage in conflict with other subreddits (e.g., \sr{The\_Donald}(8.09\%), \sr{SandersForPresident}(6.75\%), \sr{politics}(5.71\%)). However, in many cases, these values are higher than the median.

\subsubsection{Banned subreddits}
Three out of nine banned subreddits in the conflict graph rank within the top 10 when ranked by con\_author\_percent and average conflict intensity. Table~\ref{table:banned_stats} show rank (and value) by con\_author\_percent and average conflict intensity for all nine banned subreddits (lower ranks means higher con\_author\_percent and higher average intensity respectively). We observed that moderately low ranks in both measures for three other banned subreddits. Two controversial moderated subreddits \sr{Mr\_Trump} (rank 37 by con\_author\_percent and rank 20 by average intensity) and \sr{ForeverUnwanted} (rank 74 by con\_author\_percent and rank five by average intensity) also rank low when ranked by both measures. High con\_author\_percent means that a large fraction of the corresponding subreddit is participating in norm-violating behavior and high average intensity means that a large fraction of common authors between the source and target subreddits are norm-violating. Low ranks by both these measures should indicate that the corresponding subreddit is misbehaving as a community. This is supported by the fact that 6 out of 9 banned subreddits and two controversial subreddits (both set to private by moderators of the respective subreddits) display this behavior. We emphasize again that subreddits can be banned due to their content and not due to the conflict they caused. Such subreddits will not rank low in these two measures.

\begin{table}[htb] 
\centering
\begin{adjustbox}{max width=\columnwidth}
\begin{tabular}{|c|c|c|}
\hline
\textbf{Subreddit} & \textbf{Con\_author\_percent rank (value)} & \textbf{Average conflict intensity rank (value)}\\
\hline
\sr{PublicHealthWatch} &  1 (35.25) &  2 (0.62)\\
\hline
\sr{europeannationalism} &  5 (32.43) &  1 (0.75)\\
\hline
\sr{WhiteRights} &  3 (32.60) &  9 (0.34)\\
\hline
\sr{altright} &  23 (18.18) &  19 (0.24)\\
\hline
\sr{european} &  31 (17.41) &  24 (0.20)\\
\hline
\sr{Incels} &  183 (7.87) &  57 (0.11)\\
\hline
\sr{uncensorednews} &  13 (19.94) &  123 (0.06)\\
\hline
\sr{DarkNetMarkets} &  635 (1.59) &  494 (0.02)\\
\hline
\sr{SanctionedSuicide} &  475 (2.94) &  199 (0.04)\\
\hline
\end{tabular}
\end{adjustbox}
\caption{Banned subreddits and their ranks and values by average conflict intensity and con\_author\_percent.}\label{table:banned_stats} 
\end{table}

\section{Co-Conflict Communities}
\subsection{Creating the Co-conflict Graph}
Although most controversial authors have only one anti-social home, there are nonetheless patterns of conflict directed from one subreddit against multiple others. Subreddits targeted by same set of authors gives us further insight about these authors and the subreddits they call home. Using all subreddits from the conflict graph, we can create graphs that map the subreddits that are co-targeted. In the co-conflict graph, nodes are still subreddits. Edges are determined by generating a weighted edge between two subreddits $A$ and $B$ if the Jaccard coefficient between the set controversial authors, who have anti-social homes in $A$ and $B$, is positive. The Jaccard coefficient denotes how many of such authors $A$ and $B$ have in common compared to distinct negatively perceived controversial authors in both subreddits. If $X$ and $Y$ denotes the set of such authors in subreddit $A$ and $B$ respectively, the Jaccard coefficient between $X$ and $Y$ is defined as: 
\begin{center}
$Jaccard(X,Y) = \frac{X \cap Y}{X \cup Y}$
\end{center}

We also make sure that there are at least 2 common negatively perceived controversial authors between subreddits $A$ and $B$, so that we do not misidentify an edge due to one single author. 

\subsection{Co-conflict Graph Properties}
As majority of controversial authors misbehave in only one subreddit, the co-conflict graph has many disconnected components. We only focus on the largest connected component (i.e. the giant component) which consists of 237 nodes and 780 edges. Unlike the subreddit conflict graph, the co-conflict graph is undirected. Furthermore, edge semantics are different as edges denotes the similarity between two subreddits. Common network analysis algorithms can be applied to this graph more intuitively. Use of community detection, for example, can help us determine which groups of subreddits (rather than pairs)  are `co-targeted.' There are multiple algorithms for community detection in undirected networks~\cite{Fortunato2010survey} (e.g., FastGreedy, InfoMap, Label Propagation, Louvain or Multilevel, Spinglass and Walktrap). The algorithms have different trade-offs~\cite{AldecoaM13,lancichinetti_community_2009,Prat-Perez2014}, though generally both Louvain and Infomap are shown to perform well. Louvain or multilevel algorithm~\cite{blondel08,conf/isda/MeoFFP11} is based on modularity maximization, where modularity is a measure of cohesiveness of a network. An attractive property of Louvain is that it follows a hierarchical approach by first finding small, cohesive communities and then iteratively collapsing them in a hierarchical fashion. This approach on the co-attacked graph produced reasonably sized communities and the results of the community detection algorithm were very stable (i.e. do not change much on different runs). Note that, we use the weighted Louvain algorithm for this purpose. 

\subsection{Community Detection Results}
We evaluate the communities using $\mu$-score and clustering coefficient (CC). $\mu$-score is defined as fraction of edges from within the community to outside the community compared to all edges originating from the community. The clustering coefficient of a node is the fraction of connected neighbor pairs compared to all neighbor pairs. For a community, the CC is the average of clustering coefficients of all nodes in the community. In general, low $\mu$-score and high CC denotes a `good' community.  Using the weighted multilevel algorithm on the co-attacked graph we find 15 distinct communities. Table~\ref{table:cotroll_com} shows exemplar subreddits per community, size of the community, $\mu$-score and clustering coefficient for subreddits with at least 10 nodes in them.
\begin{table*}[htb] 
\begin{adjustbox}{max width=\textwidth}
\begin{tabular}{|c|p{6.5cm}|c|c|c|p{7cm}|}
\hline 
\textbf{No} & \textbf{Example subreddits} & \textbf{Size} & \textbf{$\mu$-score} & {cc} & description \\
\hline
1 & \sr{politics}, \sr{PoliticalDiscussion}, \sr{hillaryclinton}, \sr{SandersForPresident}, \sr{EnoughTrumpSpam}, \sr{Enough\_Sanders\_Spam}, \sr{AskTrumpSupporters}, \sr{SubredditDrama} & 74 & 0.33 & 0.41 & mostly politics and political discussion subreddits\\
\hline
2 & \sr{KotakuInAction}, \sr{conspiracy}, \sr{undelete}, \sr{MensRights}, , \sr{PublicFreakout}, \sr{WikiLeaks}, \sr{worldpolitics}, \sr{Political\_Revolution}, \sr{europe}, \sr{The\_Donald} & 39 & 0.20 & 0.29 & Political subreddits, controversial subreddits\\
\hline
3 & \sr{nsfw}, \sr{NSFW\_GIF}, \sr{woahdude}, \sr{cringepics}, \sr{trashy}, \sr{WatchItForThePlot} & 34 & 0.23 & 0.38 & mostly NSFW subreddits, subreddits making fun of others\\
\hline
4 & \sr{nba}, \sr{nfl}, \sr{baseball}, \sr{Patriots}, \sr{canada}, \sr{toronto}, \sr{ontario} & 16 & 0.17 & 0.15 & sports subreddits, Canada related subreddits\\
\hline
5 & \sr{TopMindsofReddit}, \sr{AgainstHateSubreddits}, \sr{worstof}, \sr{ShitAmericansSay}, \sr{SRSsucks}, \sr{TrollXChromosomes} & 15 & 0.19 & 0.24 & Subreddits focusing on other subreddits\\
\hline
6 & \sr{Overwatch}, \sr{DotA2}, \sr{GlobalOffensive}, \sr{NoMansSkyTheGame}, \sr{leagueoflegends} & 11 & 0.06 & 0.00 & video game related subreddits\\
\hline
7 & \sr{guns}, \sr{progun}, \sr{Firearms}, \sr{gunpolitics}, \sr{shitguncontrollerssay} & 10 & 0.06 & 0.23 & gun-related subreddits\\
\hline
8 & \sr{relationships}, \sr{OkCupid}, \sr{AskMen}, \sr{AskWomen}, \sr{niceguys}, \sr{instant\_regret}, \sr{sadcringe}, \sr{TheBluePill} & 10 & 0.39 & 0.51 & relationship subreddits, satirical subreddits\\
\hline

\end{tabular}
\end{adjustbox}
\caption{Communities in co-conflict network with at least 10 nodes. For each community, exemplar subreddits, size of the community, $\mu$-score and clustering coefficient(cc) is shown}\label{table:cotroll_com} 
\end{table*}

\begin{figure}[h]
 \centering
 \includegraphics[scale=0.3]{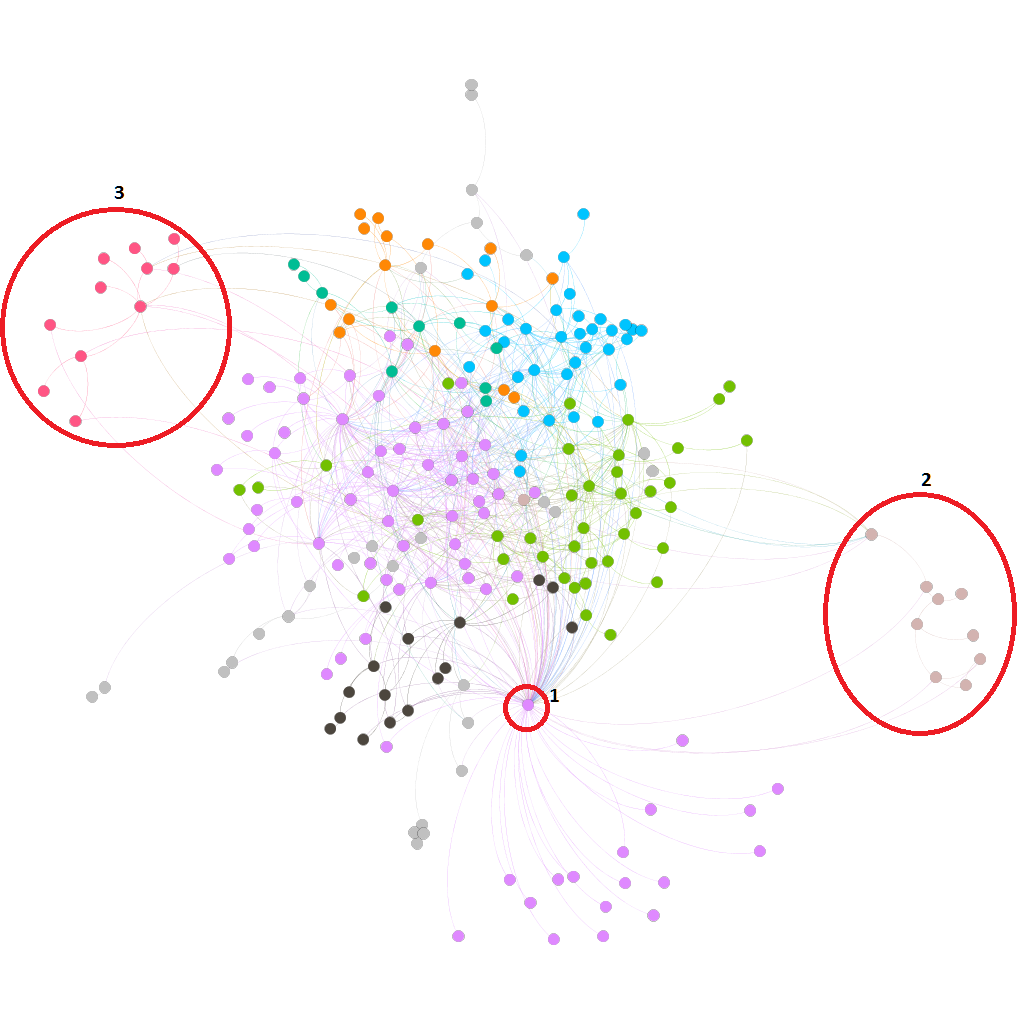}
 \caption{The Co-conflict Graph. Different communities are shown in different colors. 1 denotes is \sr{politics}, a subreddit demonstrating star pattern. 2 denotes the gun-related subreddit community and 3 denotes community of video game subreddits.}
\label{fig:co-attacked}
\end{figure}

Figure~\ref{fig:co-attacked} shows the co-conflict graph and its communities. In general, most communities show low $\mu$-score and low clustering coefficient due to presence of star-like structures (i.e. a large number of nodes are connected to one single node). For example, \sr{politics} is connected to 103 other subreddits. Smaller subreddit communities are topically more cohesive compared to larger communities. For example, community 6 (video game subreddits) and 7 (gun-related subreddits) in table~\ref{table:cotroll_com} are both topically very cohesive.

It is worth re-emphasizing that the co-conflict graph does not necessarily mean that a pair of subreddits in the same community are `friendly' and do not have a conflict with each other. For example, \sr{Christianity} and \sr{atheism} belong to same community and there are many authors who have a social home in \sr{Christianity} and anti-social home in \sr{atheism}. Similarly, \sr{SandersForPresident} and \sr{Enough\_Sanders\_Spam} are in the same community and are very much ``at war''. This is mostly due to presence of aforementioned star-like structures. For example, \sr{Republican} and \sr{democrats} both are only connected to \sr{politics} and thus belong in the community containing \sr{politics}. This does not mean that \sr{Republican} and \sr{democrats} have a common group of people perceived negatively.

\section{Conflict Dynamics}
One interesting question for our conflict graphs is how they change over time? It is possible that controversial authors maintain the same social and anti-social homes over time. Conversely, a subreddit with controversial authors may `shift' its negative behaviors to different subreddits over time. To better understand these dynamics, we study this in both an aggregate manner (i.e. does the most targeted and most instigating subreddits vary each month or do they remain mostly static?), and from the perspective of a few individual subreddits (how does rank of a particular subreddit among most targeted and most instigating subreddits vary over time?). To do so, we created conflict graphs for each month in 2016. These monthly graphs use the same set of subreddits and the same set of controversial authors used in constructing the yearly conflict graph. 

\begin{figure}[htbp!]
 \centering
 \includegraphics[scale=0.55]{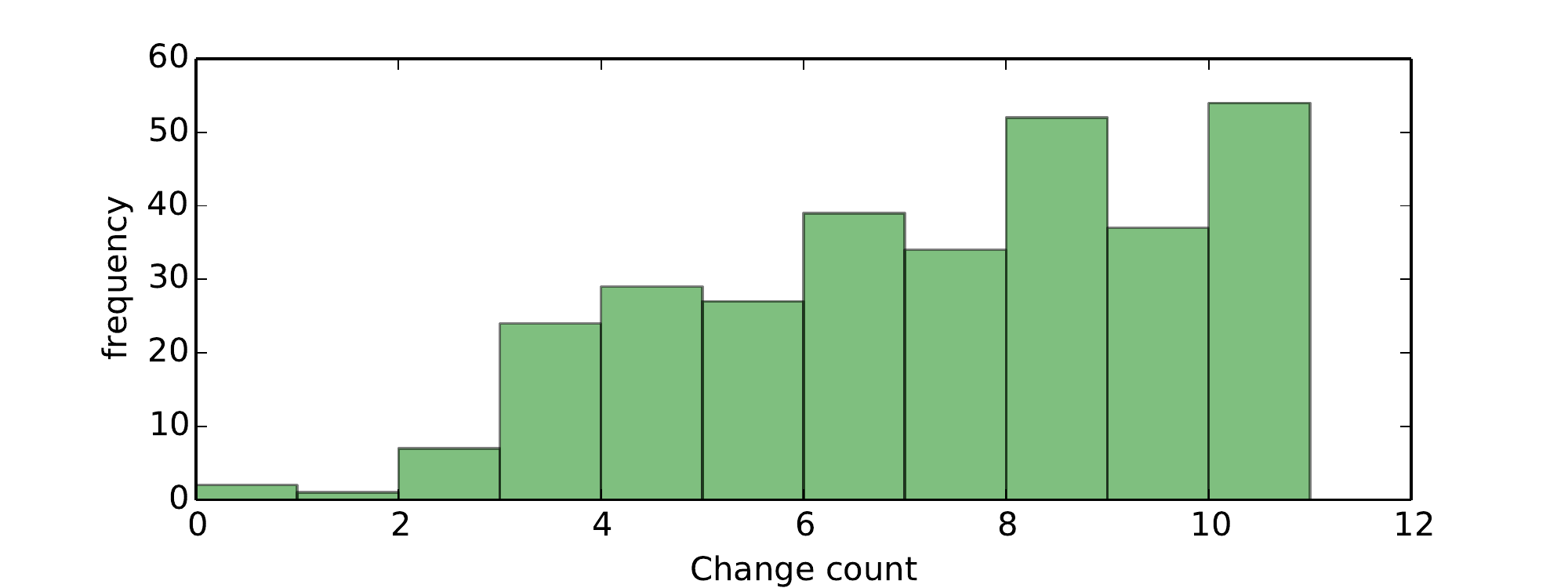}
 \caption{Change count for source subreddits who targeted at least 5 subreddits }
\label{fig:change_count}
\end{figure}

We focus this preliminary analysis on subreddits that targeted five or more other subreddits over the year and model how their `conflict focus' varies. That is, do they specifically focus on a single subreddit over all months, or does their most targeted subreddit in a specific month vary from month to month? To determine this, we count the number of times the most targeted subreddit for each conflict source subreddits change from one month to the next. We call this the \textit{change\_count} for the attacking subreddit. By definition, change\_count can vary from 0 (most targeted subreddit did not change in all 12 months) to 11 (most targeted subreddit changed every month). If a subreddit did not target any other in a particular month, but targeted some subreddit in the next (or vice versa), we count that as a change. Figure~\ref{fig:change_count} shows the distribution of change\_count for source subreddits. 

On average, we find that change\_count is 6.91 (median of 7), which means that most subreddits shifts their primary focus over time. We find only 2 subreddits did not change their target at all in 12 months. One example of this is \sr{CCW} (concealed carry weapons subreddit) targeting \sr{GunsAreCool} (a subreddit advocating for gun control in USA).

Because of the 2016 US election, the monthly `most targeted' and `most instigating' subreddits are still predominantly political. However, some subreddits only appear in the beginning of the year (e.g. \sr{SandersForPresident} is in the list of top 3 most instigating subreddits for the first four months, \sr{The\_Donald} is the top 3 most targeted subreddit list for the first 3 months) or end of the year (e.g. \sr{EnoughTrumpSpam} is in the list of top 10 most targeted subreddits for the last 7 months and during that time, it is the most targeted subreddit). On the other hand, some subreddits show remarkable consistency -- \sr{The\_Donald} is always the most instigating subreddit (for all 12 months) and \sr{politics}, \sr{SubredditDrama} are always in the top 10 most targeted subreddits list. 

\begin{comment}
\begin{figure*}
\begin{adjustbox}{max width=\textwidth}
\centering
\begin{minipage}{\columnwidth}
\centering
  \includegraphics[width=\linewidth]{figs/most_attacked_month.pdf}
  \captionof{figure}{Rank by intensity of being targeted for four political subreddits over 2016.}
  \label{fig:attacked_month}
\end{minipage}
\hspace{0.05\linewidth}
\begin{minipage}{\columnwidth}
\centering
  \includegraphics[width=\linewidth]{figs/most_attacking_month.pdf}
  \captionof{figure}{Rank by conflict intensity for four political subreddits over 2016.}
  \label{fig:attacking_month}
\end{minipage}
\end{adjustbox}
\end{figure*}
\end{comment}

\begin{figure}[htbp!]
\centering
  \includegraphics[scale=0.55]{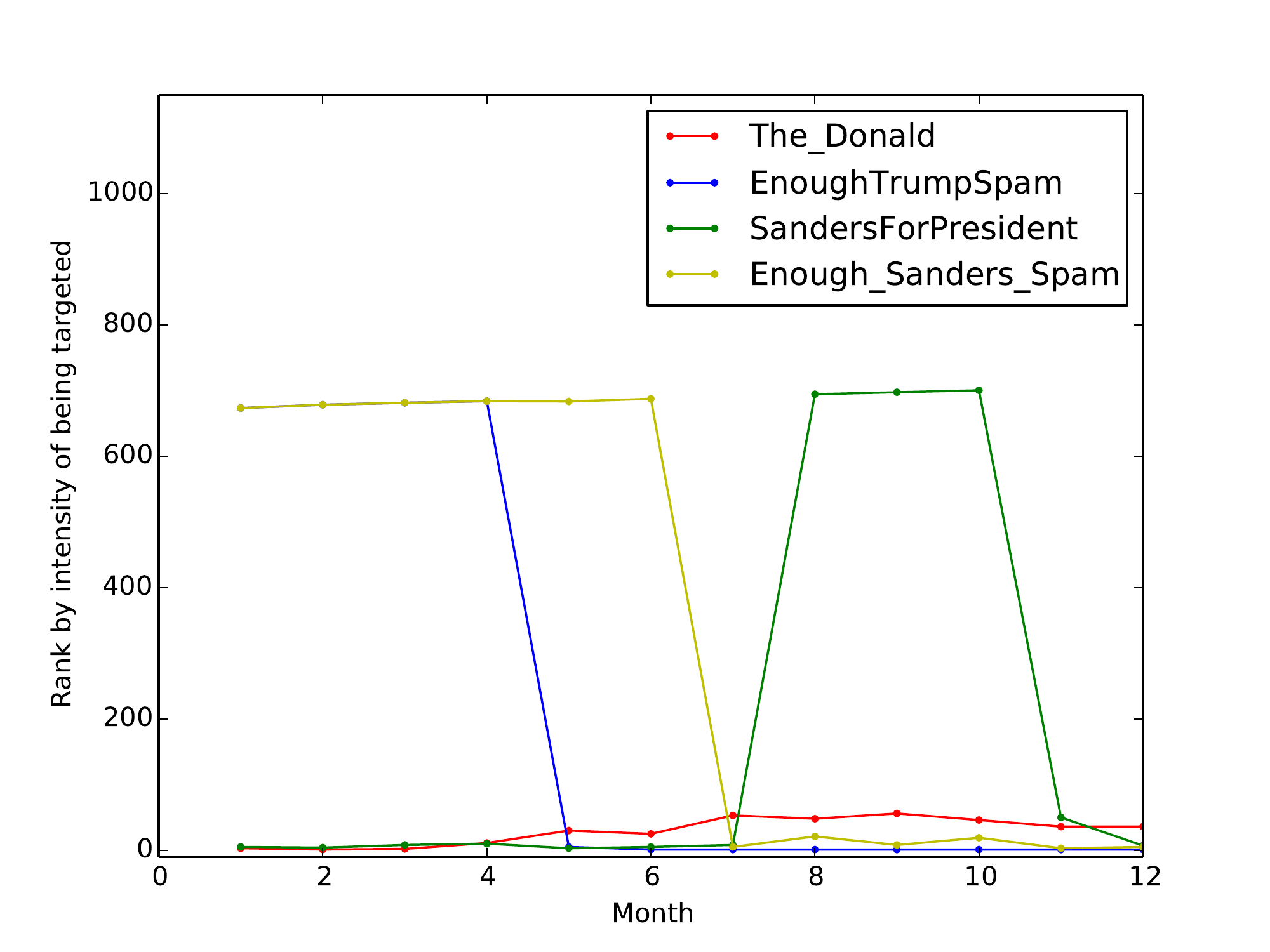}
  \captionof{figure}{Rank by intensity of being targeted for four political subreddits over 2016.}
  \label{fig:attacked_month}
\end{figure}

\begin{figure}[htbp!]
\centering
  \includegraphics[scale=0.55]{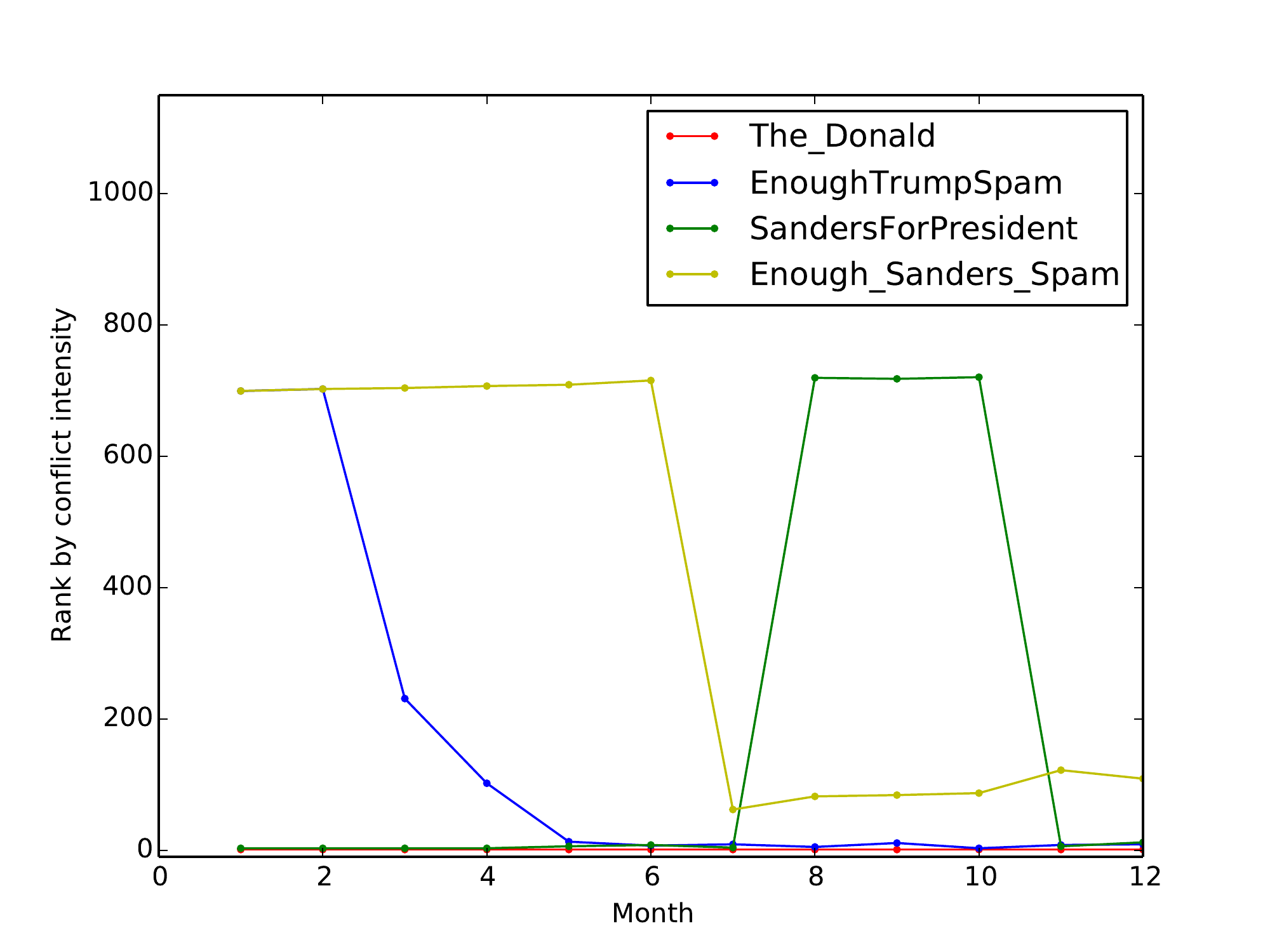}
  \captionof{figure}{Rank by conflict intensity for four political subreddits over 2016.}
  \label{fig:attacking_month}
\end{figure}

Figures~\ref{fig:attacked_month} and~\ref{fig:attacking_month} illustrate the rank of four political subreddits related to the US presidential election (\sr{The\_Donald}, \sr{EnoughTrumpSpam}, \sr{SandersForPresident} and \sr{Enough\_Sanders\_Spam}). The figures capture the rank of these in the most targeted (largest indegree in the conflict graph) and most instigating lists (largest outdegree) respectively.  These demonstrate both the pattern of stable conflict as well as varying ones.

Perhaps the most important observation from these plots is how mirrored they are. \sr{The\_Donald} is always the most instigating subreddit and it is consistently targeted back. \sr{EnoughTrumpSpam} gained popularity during March 2016 and gradually became more instigating in the next two months. For the last seven months of 2016, \sr{EnoughTrumpSpam} is the most targeted subreddit. \sr{SandersForPresident} is near the top in both most targeted and most instigating list until the end of July 2016 and from November 2016. However, in three months between July and November, this subreddit did not have any antagonistic relation with any other subreddit as it was shutdown after US political conventions in July and subsequently brought back after in November. \sr{Enough\_Sanders\_Spam} was formed in July 2016 and instantly became highly targeted due to its content. This shows that, a subreddit instigating/targeted can be highly dependant on external events.

\section{Discussion}
In this paper, we demonstrate a quantitative method for identifying community-to-community conflicts by aggregating users who behave differently depending on the community they interact with. We define social and anti-social homes of a user based on a local perception of norm-compliance and norm-violation (which we measure by reward and sanction through voting). This method allows us to find conflict in any social network with `noisy' community structure. Though we focus on Reddit in a specific year (2016), we believe the work is more broadly usable both across time and other social media sites. 

Before discussing in which situations our approach may or may not be usable, we briefly summarize our key findings. We find that community-to-community conflicts are usually reciprocal but mutual conflict intensities usually do not match up. We identify which subreddits generated most conflict and which subreddits were most targeted. By analyzing subreddits banned by Reddit in relation to our measures  (e.g., average conflict intensity, a high percentage of positively perceived controversial authors, etc.) we illustrate how our technique may be useful for identifying problematic subreddits. Co-conflict subreddit communities show that subreddit conflicts are not random in nature, as we observe topically similar subreddits usually belong to the same co-conflict community. We perform a preliminary analysis of temporal patterns in subreddit conflicts and find that the conflict focus usually shifts over time. 

Below we focus on the generalizability and limitations of our findings and approach. Specifically, we discuss the appropriateness and alternatives to using up/down-votes to determine conflicts, contrast of a subreddit conflict with topically opposite subreddits, robustness of the threshold parameters, the potential of communal misbehavior versus behavior of only a few members of a given community and the co-conflict graph. 

\subsection{Downvotes for determining community conflicts}
A downvoted comment in a particular subreddit may be a reaction to a number of factors ranging from innocuous norm-violation, being off-topic, presenting a non-conforming viewpoint, low-effort posts (e.g. memes), reposts, and truly malicious behavior. Furthermore, social news aggregation websites like Reddit generally skew towards positive feedback, and are susceptible to social influence effects~\cite{Muchnik647}. Because any individual comment, or even author, may receive up- or down-votes due to these factors, we rely on aggregate signals in our analysis.  Thus, a user having many comments with downvotes (these comments may have more upvotes compared to downvotes) or downvoted comments at all, might point to existence of an anti-social home of the user. However, we opt for a more stringent definition of anti-social home due to two reasons. First, Reddit provides a comment score which is simply the number of upvotes minus the number of downvotes, but does not provide the exact number of up and downvotes as a measure of reducing spam-bot activity ~\footnote{\url{https://www.reddit.com/wiki/faq}}. Thus, we can not use the number of up and downvotes of a comment provided by Reddit as a reliable metric. Furthermore, a user in a subreddit might have downvotes (and a few downvoted comments) due to being new in the subreddit (i.e., not knowing all the rules) or brigading, where users from antagonistic subreddits downvote random or targeted comments as a `downvote brigade'. However, we would like to point out that our definition and threshold parameters of social and anti-social homes are not set in stone and can be easily adapted for similar definitions or threshold parameters, without changing the rest of the algorithmic pipeline to determine the conflict graph.

It is also worth repeating that a downvote does not provide a \textit{global} quality assessment of a comment. Rather, a downvoted comment within a subreddit signifies that this particular subreddit perceives the comment as low quality. This is a localized definition of quality defined by the subreddit and it is consistent with Brunton's model of spam~\cite{brunton2012constitutive}. Globally, these comments might not be seen as norm-violating or low-quality. We acknowledge the fact that users may receive negative feedback not for their own antisocial behavior, but for the antagonistic stance of the receiving community. We do not assume that, for a conflict edge, the instigating community is a `community of aggressors'. In fact, depending on the viewpoint, it might be viewed as a `refuge for social outcasts.' New users in a subreddit are more susceptible to innocuous norm-violation due to them not knowing all rules of a new subreddit, but with time they tend to learn. To eliminate these users from the list of controversial users, we enforce a minimum threshold of comments in a subreddit. Excluding these users, we use downvote within a subreddit to determine subreddit conflicts and not as an indicator of the global quality of the comment.

\subsection{Subreddit conflict due to ideological differences}
Many subreddit conflicts in the subreddit conflict graph are between subreddits with topically or ideologically opposing viewpoints. This is expected given the high presence of political subreddits in the graph. However, identifying subreddits with ideologically differing viewpoints does not signify a subreddit conflict. Similarly, topical differences do not explain all subreddit conflicts. Two subreddits with opposing ideologies may engage in a civil discussion about the topic, may not engage at all or be part of a subreddit conflict. In many cases, engagement is very low or non-existent. An example of this is \sr{askscience} and \sr{theworldisflat}~\cite{Datta2017z2}. Similarly, the conflict edge from \sr{The\_Donald} to \sr{PanicHistory} can not be fully explained by topical opposition. On the other hand, the presence of subreddit conflicts between many ideologically antagonistic subreddits works as a sanity check and provide insight into what kind of ideological opponents are more likely to engage in community level conflicts.

\subsection{Robustness of threshold parameters}
We employ multiple thresholds to ensure proper conflict identification. Although some of the thresholding can be eliminated via other methods~\cite{Martincommunity2vec}, we use this approach for computational simplicity and effectiveness. Nonetheless, threshold parameters must still be tuned for the particular dataset and application. We discuss our philosophy behind choosing different parameters and justify our choice via a set of small-scale sensitivity analyses. We focus on key differences for different threshold values.

The first threshold is the number of comments by a user. We only consider users who commented more than 100 times to ensure that we perform our analysis on active users. However, we find that our results are quite robust to change in the threshold. Considering users who commented more than 50 times we see around 10\% increase in the number of controversial authors. On the other hand, a stricter threshold of 200 reduces the number by 16\%. The conflict graphs generated using these thresholds also show little change. We observe 3.1\% increase in conflict graph nodes and 1.1\% increase in conflict edges using threshold 50, and 4.2\% decrease in nodes and 2.7\% decrease in edges using a threshold 200. Removing overall low activity accounts from consideration removes malicious sockpuppet accounts (i.e., a single user uses multiple accounts usually unlinked with each other). Unfortunately, we can not directly account for these users as we do not have the data. However, with knowledge of sockpuppets, we can merge multiple accounts before thresholding, which retains the behavior of the sockpuppet account in the aggregate.

The next major threshold we use is determining the minimum number of comments for a user to have a social or anti-social home. We settled for users having more than 10 comments in a subreddit. This threshold works as a trade-off between adding genuine social and anti-social homes for users with lower activity, possibly in smaller subreddits and falsely identifying social and anti-social homes due to low user activity. Using a lower threshold of more than five comments adds 131\% more controversial authors, which in turn adds 124\% more nodes and 378\% more edges in the conflict graph. On the other hand, using a stricter threshold of more than 20 comments, eliminated 63\% controversial authors and 62\% nodes and 86\% edges in the conflict graph. This stricter threshold also eliminates five of nine banned subreddits from the conflict just because they are not very large in size. It can be argued that different thresholds should be used for social and anti-social homes as one-off malicious comments from many users can overwhelm subreddits. However, due to ease of account creation in Reddit, these one-off comments are often done via `throwaway accounts' created for the explicit purpose of anti-social behavior and it is difficult to link these sockpuppets to specific communities. Moreover, it is very difficult to identify truly malicious one-off comments from innocuous norm-violations or low-effort posting as we do not have labeled data and deciphering the true intention behind these comments are often context-sensitive (i.e. same comment can be perceived as malicious or non-malicious depending on the context). Our experience is lowering the threshold for determining anti-social homes add many false conflict edges. We choose to err on the side of caution by having a somewhat strict threshold without eliminating most smaller low activity subreddits. We acknowledge that while this approach finds social and anti-social home for long-term misbehaving users, it does not capture sudden conflicts risen from strong external stimuli (e.g., \textit{2015 AMAgeddon}) or when long-standing contributors to a community suddenly starts `misbehaving'~\cite{hardaker2010trolling}.

The thresholds for determining conflict edges (at least five controversial authors behaving differently in a pair of subreddits) and co-conflict graph edges (at least two controversial authors perceived negatively in a pair of subreddits) are somewhat lenient, as our definition of a controversial author is quite strict. We observe, a stricter threshold in both above-mentioned cases, eliminates smaller low-activity subreddits from consideration. 

\subsection{Identifying communal misbehavior}
Many conflict edges in the conflict graph have low intensity and most subreddits have low con\_author\_percent value. In other words, only a few individuals in a subreddits compared to subreddit size are controversial authors. We can infer that getting involved in subreddit conflicts does not imply communal misbehavior. In fact, larger subreddits are more likely to be involved in more conflicts due to their size. However, there is an important distinction in a conflict edge compared to a few pathological individuals behaving badly. We determine conflicts via controversial authors, which means the users who are perceived negatively in the target subreddit are perceived positively in the source subreddit. This implies either these controversial authors behave very differently in the source compared to the target, or users in the source subreddit support the controversial author's behavior. We do not look for such distinctions in this paper. However, this distinction may be useful to isolate in future work.  
To determine community-wide misbehavior we primarily look into what percentage of active subreddit members are positively perceived controversial authors (con\_author\_percent). For many banned subreddits (where we infer communal misbehavior because these were banned) we observe that more than 30\% of subreddit users fall into this category. Many banned subreddits also show high average outgoing conflict intensity. In general, Reddit users restrict themselves to posting in only a few subreddits~\cite{Hamilton2017}. Thus, a conflict edge with high intensity shows that whatever little interaction the participating subreddits have, is toxic. Notably, due to high variance of subreddit sizes, only a few people from a large subreddit can potentially overwhelm a smaller subreddit even if the number of misbehaving users is very low compared to the size of the larger subreddit. We believe that both high con\_author\_percent and high average outgoing conflict intensity implies communal misbehavior. We would emphasize that this is not the only way a community can misbehave. Abusive language, anti-social or unlawful behavior within the subreddit can also point to communal misbehavior and can lead to subreddit bans.     

Moreover, we would like to encourage discussion about communal behavior versus behavior of only a few members in the community in a general sense. It is not always clear what threshold one should abide by when declaring a particular behavior as `communal' (e.g., what percentage of community members must behave in a certain way to consider that behavior as communal). This discussion applies to Reddit as well as many other online social platforms which exhibit community patterns. We believe that the con\_author\_percent measure for banned subreddits can be used as a starting point of identifying community-wide misbehavior at least for different subreddits.

\subsection{Co-conflict graph}
The co-conflict graph embodies anti-social home to anti-social home relationships among the same set of subreddits as the conflict graph. As 82\% of the controversial authors have only a single anti-social home, the co-conflict graph is sparser compared to the conflict graph. Communities in the co-conflict graph identify which meta-subreddit groups are targeted together. As one might expect, some of the co-conflict graph communities are extremely topic-coherent (gun-related and video-game subreddits). However, other groupings provide additional insights. For example, a larger subreddit, when targeted together with many comparatively smaller subreddits, forms a star pattern. The meta-subreddit groups are not necessarily targeted by another meta-subreddit groups as we observe a lot of conflicts are generated within the co-conflict subreddit communities. Interestingly, a social home to social home relationship graph form a very dense structure and does not exhibit community behavior, which means that we can not readily classify conflict between different subreddit communities using this method. 

\subsection{Limitations}
Using controversial authors to find subreddit conflicts has some limitations. First, this method does not take into account comments deleted by users or moderators (this data is not available for collection). Some subreddits are especially aggressive in deleting downvoted or moderated comments. In some cases, misbehaving authors in a subreddit are banned from further posts. As with comments, we do not have records of this type of moderation. When a subreddit aggressively bans many people, it can change the conflict graph from a static and dynamic perspective. 

A clear example of this is \sr{The\_Donald}, which banned thousands of individuals over its lifetime (these banned individuals formed a subreddit \sr{BannedFromThe\_Donald}, with a subscriber count of 2,209 in November of 2016 and over 27,000 in July of 2018). These individuals do not show up as controversial authors as their comments are gone. We also do not account for sockpuppetry, i.e., having multiple accounts, one for normal posting behavior on Reddit and others for misbehaving. Presence of many users with sockpuppets can skew the estimation of controversial authors in different subreddits. 

If data such as bans on the source of sockpuppet accounts can be determined, this data could easily be incorporated in our algorithmic pipeline by updating the definition of anti-social homes. For example, if we know the users who are banned from a particular subreddit, we declare that these users have an anti-social home in the subreddit they are banned from. 

In our current analysis, we do not filter bots (software applications that generate comments) from our list of authors. However, strictly malicious bots -- those with \textit{only} anti-social homes -- do not change our conflict graph as they are not counted among the controversial authors. Occasionally, bots can show up as controversial authors. These include moderator bots (e.g., \textit{AutoModerator}). It is worth a future study to understand why a bot can be perceived positively or negatively depending on the subreddit. This might mean that bots can intentionally, or not, violate norms for some subreddits while complying with others. Though bots represent a small fraction of Reddit users, this behavior would be interesting for future study.

One final limitation of our model is that \textit{correlated} multi-community posting may appear as a conflict edge.  For example, members of community $A$ (a subreddit for a specific computer game) are found to conflict with community $B$ (a feminist subreddit). However, it may not be appropriate to say that $A$ conflicts with $B$. The topics of the two communities are completely orthogonal. In this situation, it might be due to the presence of a third subreddit, $C$ (e.g., an anti-feminist subreddit) that conflicts with $B$. It simply happens that many members of $A$ (the game) also have a social-home on $C$ (the anti-feminist subreddit). It would thus be more accurate to say that $C$ conflicts with $B$. One approach for handling this is to ensure that there is some topical correspondence between the communities we are considering (based on text). This eliminates the $A-B$ edge but retains $C-B$. It is nonetheless possible that we may want to know that the $A-B$ link exists. Moderators of subreddit $A$ might want to be made aware of this correlation and take action. 

\subsection{Implications}
Although we perform our analysis on Reddit, our analysis is equally applicable in other social media with inherent or inferred community structure with associated community feedback. For example, we can perform a similar analysis on Facebook pages and groups, online news communities and Twitter hashtag communities (people who tweeted a particular hashtag are part of that hashtag community). We quantify user behavior based on upvotes and downvotes in a particular community, and this data is more easily available for many social media websites compared to a list of banned or otherwise sanctioned users from a particular community. Our approach is highly adaptable and can incorporate new information (e.g., banned and sockpuppet accounts). The analysis is also fully automated and highly parallelizable which increases the adaptability for a very large amount of data.    

In addition to providing insight into communities, we also believe that our work can be used for moderation purposes. We observe that several banned subreddits rank very high on particular metrics for measuring conflict. We can calculate these measures for monthly (or otherwise temporal) subreddit conflict graphs and see how different subreddits rank in these measures over time. This observation can be used to monitor problematic subreddit behavior as a whole or create an early-warning system based on machine learning where we treat currently banned subreddits as positive examples of communal misbehavior and use the metrics above as features.
\section{Conclusion}
In this work, we study community-on-community conflict. We describe a mechanism for determining the social and anti-social homes for authors based on commenting behavior. From these, we construct `conflict edges' to map the conflicts on Reddit. Using our approach, we allow for a contextual definition of anti-social behavior based on local subreddit behavior. This provides a different perspective than studying global-norm violating behaviors.

We find that most conflicts (77.2\%) are reciprocated, but the intensities from both sides do not necessarily match up. Larger subreddits are more likely to be involved in more subreddit conflicts due to their large user-base, but most of these conflicts are minor, and this does not imply large-scale communal misbehavior. On the other hand, we find that high average conflict intensity and a large fraction of subreddit users perceived negatively in other subreddits may have implications for communal misbehavior. Finally, we explore temporal patterns in conflicts and find that subreddits that target multiple others, will shift their main conflict focus over time. We believe this analysis can be applied to other social media sites which display community structure, create early warning systems for norm-violating communities and help encourage discussion about community-wide misbehavior in social media.    
\section{Acknowledgement}
We'd like to thank Chanda Phelan and David Jurgens for their valuable feedback.

% REFERENCES FORMAT
% References must be the same font size as other body text.
\bibliographystyle{plain}
\bibliography{references} 

\end{document}